\documentclass{elsart}
\usepackage{graphicx,amssymb}
\journal{Advances in Space Research}
\begin{document}
%
%%%%%%%%%%%%%%%%%%%%%%%%%%%%%%%%%%%%%%%%%%%%%%%%%%%%%%%%%%%%%%%%%%%
\begin{frontmatter}
\title{
The energy spectrum of all-particle \\ cosmic rays around the 
knee region \\ observed with the Tibet-III air-shower array
}
\author[1]{M.Amenomori},
\author[2]{S.Ayabe},
\author[3]{X.J.Bi},
\author[4]{D.Chen\corauthref{cor}},
\corauth[cor]{Corresponding author.}
\ead{chen@icrr.u-tokyo.ac.jp}
\author[5]{S.W.Cui},
\author[6]{Danzengluobu},
\author[3]{L.K.Ding},
\author[6]{X.H.Ding},
\author[7]{C.F.Feng},
\author[3]{Zhaoyang.Feng},
\author[8]{Z.Y.Feng},
\author[9]{X.Y.Gao},
\author[9]{Q.X.Geng},
\author[6]{H.W.Guo},
\author[3]{H.H.He},
\author[7]{M.He},
\author[10]{K.Hibino},
\author[11]{N.Hotta},
\author[6]{Haibing Hu},
\author[3]{H.B.Hu},
\author[12]{J.Huang},
\author[8]{Q.Huang},
\author[8]{H.Y.Jia},
\author[13]{F.Kajino},
\author[14]{K.Kasahara},
\author[4]{Y.Katayose},
\author[15]{C.Kato},
\author[12]{K.Kawata},
\author[6]{Labaciren},
\author[16]{G.M.Le},
\author[7]{A.F.Li},
\author[7]{J.Y.Li},
\author[17]{Y.-Q.Lou},
\author[3]{H.Lu},
\author[3]{S.L.Lu},
\author[6]{X.R.Meng},
\author[2,18]{K.Mizutani},
\author[9]{J.Mu},
\author[15]{K.Munakata},
\author[19]{A.Nagai}
\author[1]{H.Nanjo},
\author[20]{M.Nishizawa},
\author[12]{M.Ohnishi},
\author[21]{I.Ohta},
\author[2]{H.Onuma},
\author[10]{T.Ouchi},
\author[12]{S.Ozawa},
\author[3]{J.R.Ren},
\author[22]{T.Saito},
\author[12]{T.Y.Saito},
\author[13]{M.Sakata},
\author[12]{T.K.Sako},
\author[10]{T.Sasaki},
\author[4]{M.Shibata},
\author[12]{A.Shiomi},
\author[10]{T.Shirai},
\author[23]{H.Sugimoto},
\author[12]{M.Takita},
\author[3]{Y.H.Tan},
\author[10]{N.Tateyama},
\author[17]{S.Torii},
\author[24]{H.Tsuchiya},
\author[12]{S.Udo},
\author[9]{B.Wang},
\author[3]{H.Wang},
\author[12]{X.Wang},
\author[7]{Y.G.Wang},
\author[3]{H.R.Wu},
\author[7]{L.Xue},
\author[13]{Y.Yamamoto},
\author[12]{C.T.Yan},
\author[9]{X.C.Yang},
\author[25]{S.Yasue},
\author[16]{Z.H.Ye},
\author[8]{G.C.Yu},
\author[6]{A.F.Yuan},
\author[10]{T.Yuda},
\author[3]{H.M.Zhang},
\author[3]{J.L.Zhang},
\author[7]{N.J.Zhang},
\author[7]{X.Y.Zhang},
\author[3]{Y.Zhang},
\author[3]{Yi.Zhang},
\author[6]{Zhaxisangzhu},
\author[8]{X.X.Zhou}, \\
(The Tibet AS$\gamma$ Collaboration)
\address[1]{Department of Physics, Hirosaki University, Hirosaki 036-8561, Japan}
\address[2]{Department of Physics, Saitama University, Saitama 338-8570, Japan}
\address[3]{Key Laboratory of Particle Astrophysics, Institute of High Energy 
Physics, Chinese Academy of Sciences, Beijing 100049, China}
\address[4]{Faculty of Engineering, Yokohama National University, Yokohama 240-8501, Japan}
\address[5]{Department of Physics, Hebei Normal University, Shijiazhuang 050016 , China}
\address[6]{Department of Mathematics and Physics, Tibet University, Lhasa 850000, China}
\address[7]{Department of Physics, Shandong University, Jinan 250100, China}
\address[8]{Institute of Modern Physics, South West Jiaotong University, Chengdu 610031, China}
\address[9]{Department of Physics, Yunnan University, Kunming 650091, China}
\address[10]{Faculty of Engineering, Kanagawa University, Yokohama 221-8686, Japan}
\address[11]{Faculty of Education, Utsunomiya University, Utsunomiya 321-8505, Japan}
\address[12]{Institute for Cosmic Ray Research, University of Tokyo, Kashiwa 277-8582, Japan }
\address[13]{Department of Physics, Konan University, Kobe 658-8501, Japan}
\address[14]{Faculty of Systems Engineering, Shibaura Institute of Technology, Saitama 337-8570, Japan}
\address[15]{Department of Physics, Shinshu University, Matsumoto 390-8621, Japan}
\address[16]{Center of Space Science and Application Research, Chinese Academy of Sciences, Beijing 100080, China}
\address[17]{Physics Department and Tsinghua Center for Astrophysics,
Tsinghua University, Beijing 100084, China}
\address[18]{Advanced Research Institute for Science and Engineering, 
Waseda University, Tokyo 169-8555, Japan}
\address[19]{Advanced Media Network Center, Utsunomiya University, Utsunomiya 321-8585, Japan}
\address[20]{National Institute of Informatics, Tokyo 101-8430, Japan}
\address[21]{Tochigi Study Center, University of the Air, Utsunomiya 321-0943, Japan}
\address[22]{Tokyo Metropolitan College of Industrial Technology, Tokyo 116-8523, Japan}
\address[23]{Shonan Institute of Technology, Fujisawa 251-8511, Japan}
\address[24]{RIKEN, Wako 351-0198, Japan}
\address[25]{School of General Education, Shinshu University, 
Matsumoto 390-8621, Japan}
\begin{abstract}
We have already reported the first result on the all-particle spectrum 
around the knee region based on data from 2000 November to 2001 October 
observed by the Tibet-III air-shower array. In this paper,
we present an updated result using data set collected in the period 
from 2000 November through 2004 October in a wide range over 3 decades 
between $10^{14}$ eV  and $10^{17}$ eV, in which the position of the knee 
is clearly seen at around 4 PeV. The spectral index is -2.68 $\pm$ 0.02(stat.) 
below 1PeV, while it is  -3.12 $\pm$ 0.01(stat.) above 4 PeV in the 
case of QGSJET+HD model, and various systematic errors are under study now.

\end{abstract}
\end{frontmatter}
%%%%%%%%%%%%%%%%%%%%%%%%%%%%%%%%%%%%%%%%%%%%%%%%%%%%%%%%%%%%%%%%%%%
\section{Introduction}

The energy spectrum of primary cosmic rays is well described 
by a power law over a wide energy range, while its slope becomes steeper 
in the energy range between $10^{15}$ and $10^{16}$ eV, which is
called "knee". It has been discussed that the knee is closely related with
the origin, the acceleration and the propagation of high-energy cosmic rays 
in the Galaxy. One of the plausible understanding may be that
almost all cosmic rays below the knee are accelerated at supernova 
remnants (SNRs), since the maximum energy gained by shock acceleration 
at SNRs is of the order of $10^{14}$ eV per unit charge \cite{Lagage}, 
the cosmic-ray spectrum is expected to become steeper  at energies around 
and beyond the knee. Another possibility is that the break of the spectrum 
around the knee represents the energy at which cosmic rays can escape 
more freely from the trapping zone in the Galactic disk \cite{Peters}.

We have already reported the first result on the all-particle spectrum 
around the knee region based on data from 2000 November to 2001 October 
observed by the Tibet-III air-shower array \cite{Amenomori4}. In this paper, 
we present an updated result using data set collected in the period 
from 2000 November through 2004 October. We also examine the simulation 
code CORSIKA with interaction models of QGSJET01c. Based on these 
calculations, we obtained the cosmic ray energy spectrum in a wide range 
over 3 decades between $10^{14}$ eV  and $10^{17}$ eV.

%%%%%%%%%%%%%%%%%%%%%%%%%%%%%%%%%%%%%%%%%%%%%%%%%%%%%%%%%%%%%%%%%%%%%%
\section{Tibet-III air shower array and detector response}

The Tibet-III air-shower array, consisting of 533 scintillation
detectors of each 0.5 m$^{2}$ with the area of 22,050 m$^{2}$,
has been successfully operating since 1999 with energy threshold 
of a few TeV. A 0.5 cm-thick lead plate is put on the top of
each detector to increase the detection sensitivity of a detector
by converting secondary gamma rays into electron-positron
pairs in an air shower. In the fall of 2000, this array was further 
improved for UHE cosmic-ray study by adding wide dynamic range PMTs to all
221 sets of detectors which are placed on a lattice of 15 m spacing
in the detector covering area of 36,900 m$^{2}$. This PMT equipped in each
detector can measure the number of particles beyond 4,000,
so that the array can observe UHE cosmic rays with the energy exceeding $10^{17}$ eV
 with a good accuracy.

The Tibet-III air-shower array is able to measure the shower size and 
the arrival direction of each air shower. 
The air shower direction can be estimated with an inaccuracy smaller than
0.2$^{\circ}$ at energies above 10$^{14}$ eV. The shower size $N_e$ for each event
is estimated by fitting the lateral particle density distribution with the modified 
NKG structure function (see section ~\ref{Rec}).  
The primary energy of each air shower event is then estimated from the air shower size.
The relation between the shower size and the primary energy is calculated based
on the MC simulation as discussed later. In general, the ^^ ^^ shower size'' means 
 the number of electrons and positrons in each air shower event. It cannot be, however, 
directly obtained by usual shower array using scintillators since these electromagnetic 
components cannot be discriminated from other charged particles such as muons and hadrons.
 
\begin{figure}[t]
\begin{center}
\includegraphics*[width=7.5cm]{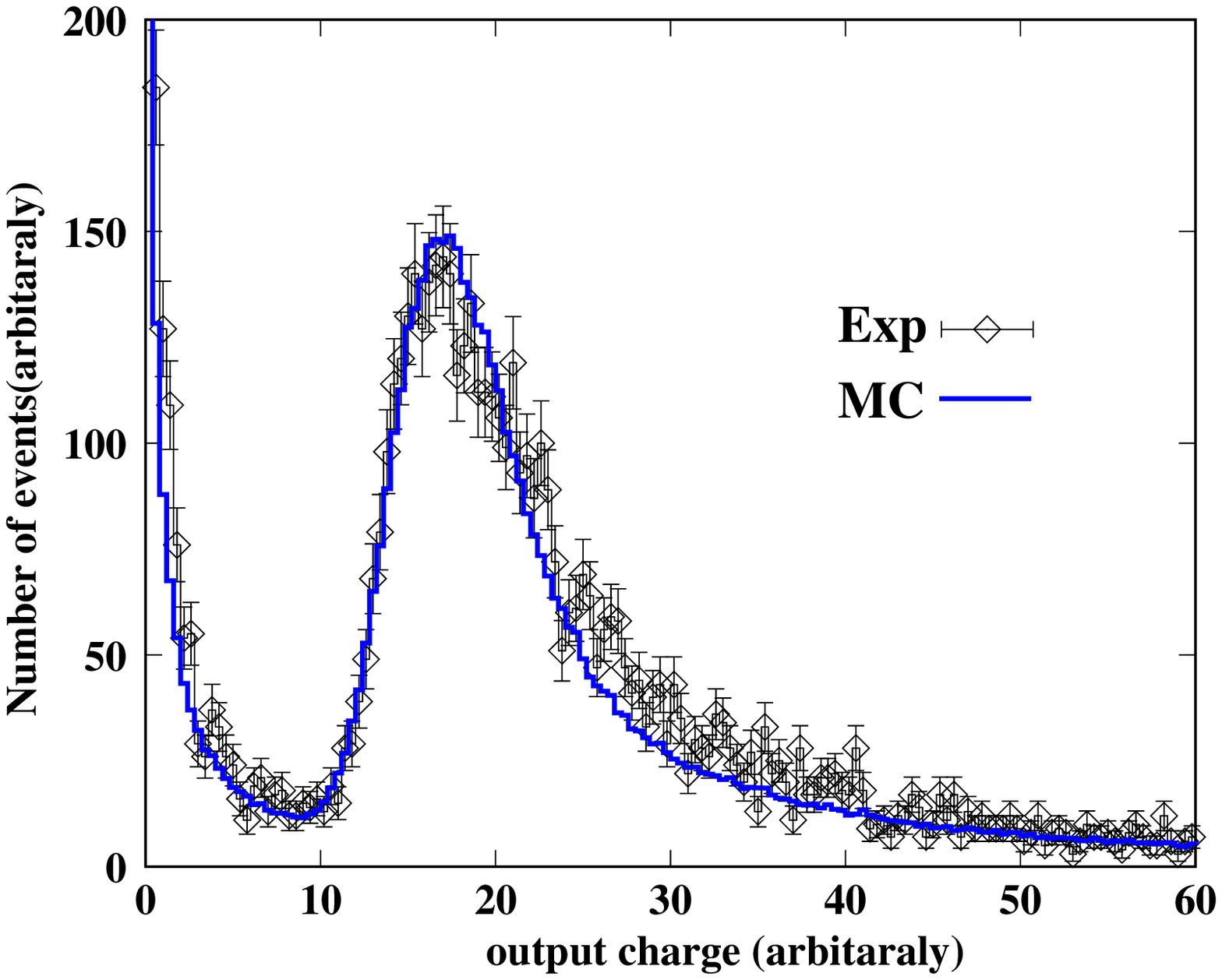}
\end{center}
\caption{Distribution of the output charge of PMT by probe calibration of 
the Tibet-III detector. In order to compare with the simulation on 
the energy deposit, the MC result is adjusted by multiplying a
constant to meet with the same peak position as the experiment.
The fluctuation of the number of photons in scintillation light is 
taken into account with the normal distribution in MC.
}
\label{fig:1}
\end{figure}

In our experiment, the number of particles detected by each  scintillation
detector is defined as the PMT output (charge) divided by that of the single peak,
which is determined by a probe calibration using cosmic rays. 
For this purpose, a small scintillator of 25 cm $\times$ 25 cm $\times$ 3.5 cm
thick with a PMT (H1949) is put on the top of the each detector
during the maintenance period. This is called probe detector and is used
for making the trigger of the each Tibet-III detector.
These events triggered by the probe detector was also examined by
a MC simulation. In this simulation, the primary particles were sampled in the
energy range above the geomagnetic cutoff energy at Yangbajing, and
 all secondary particles which pass through the probe detector and the Tibet-III detector 
were selected for the analysis. Since the value of PMT output is proportional to the 
energy loss of the particles passing through the scintillator, the peak position 
of the energy loss distribution  is defined as the ^^ ^^ energy deposit by a 
singly charged particle''. This value  corresponds  to the observed single peak 
of the probe calibration. The peak energy was calculated as 6.11 MeV for 
the Tibet-III detector configuration. We confirmed that the shape of 
the energy loss distribution shows a good agreement with the charge distribution 
of the experimental data as shown in Fig.~\ref{fig:1}. It should be noted 
that the dependences of the energy loss on particle type and its energy are 
adequately taken into account in this calibration.

%
% Fig.1
%
%

For air-shower MC calculation at high energies, we treat the number of shower
particles as the calculated energy deposit divided by 6.11 MeV.
Thus, all detector responses including muons and the materialization of photons
inside the detector are taken into account. The shower size of each event
was estimated using a modified NKG lateral distribution function which is tuned
to reproduce the above defined lateral distribution using the MC
 events under our detector configurations.

%%%%%%%%%%%%%%%%%%%%%%%%%%%%%%%%%%%%%%%%%%%%%%%%%%%%%%%%%%%%%%%%%%%%%%%%%%%%%%%%%%
\section{Air shower size and primary particle energy}
\label{SIM}

An extensive Monte Carlo simulation (MC) using CORSIKA code (ver. 6.204) 
including QGSJET01c \cite{Heck1} hadronic interaction model was made to 
 obtain the primary cosmic ray spectrum using the Tibet-III air shower data.
 Since the Tibet hybrid experiment of the air shower array and the burst 
detector array to measure the energy spectrum 
of the light components (proton and helium) strongly suggests  that the knee 
region is dominated by heavy components \cite{Amenomori5}, in this paper 
we use a heavy dominant (HD) composition model \cite{Amenomori2} in the MC 
for comparison with experimental data. All secondary particles are traced until 
their energies become 1 MeV in the atmosphere. Simulated air-shower events were 
input to the detector with the same detector configuration as the Tibet-III 
array with use of Epics code (ver. 8.64) \cite{Kasahara1} to calculate the energy 
deposit of these shower particles.

%\section{Analysis}

\subsection{Event selection and collecting area}

Following criteria are applied to select the events for
the analysis. 

1) The zenith angle ($\theta$) of each air-shower event should be smaller
than 25$^{\circ}$, or $\sec\theta$ $\leq$ 1.1 to exclude
the zenith angle dependence of the air-shower development.

2) More than 10 detectors should detect a signal of more than five 
particles per detector.

3) The central positions weighted by the 8th power of the number of 
particles at each detector should be inside the innermost 
135 m $\times$ 135 m area. This area is chosen with use of
MC events so that the following two cases are just canceling
out each other. Namely, the number of events originally inside
of this area falling outside of this area after event reconstruction 
equals to the number of events in the opposite case.

It is confirmed by simulations that the air showers
induced by primary particles with $E_0$ $\geq$ 100 TeV 
and $\sec\theta$ $\leq$ 1.1 can be fully detected without any bias under above
mentioned criteria.  

The total effective area S $\times$ $\Omega$
is calculated to be 10410.1 m$^{2}\cdot$sr for all primary particles
with $E_0$ $\geq$ 100 TeV. For the operation period from
2000 November through 2004 October, the effective live time
T is 2.21 years. The total number of air showers selected
under the above conditions is 4.1 $\times$ $10^{7}$ events. 

\subsection{Lateral distribution of shower particles}
\label{Rec}

\begin{figure}[t]
\begin{center}
\begin{minipage}[b]{8.0cm}
\includegraphics*[width=7.5cm]{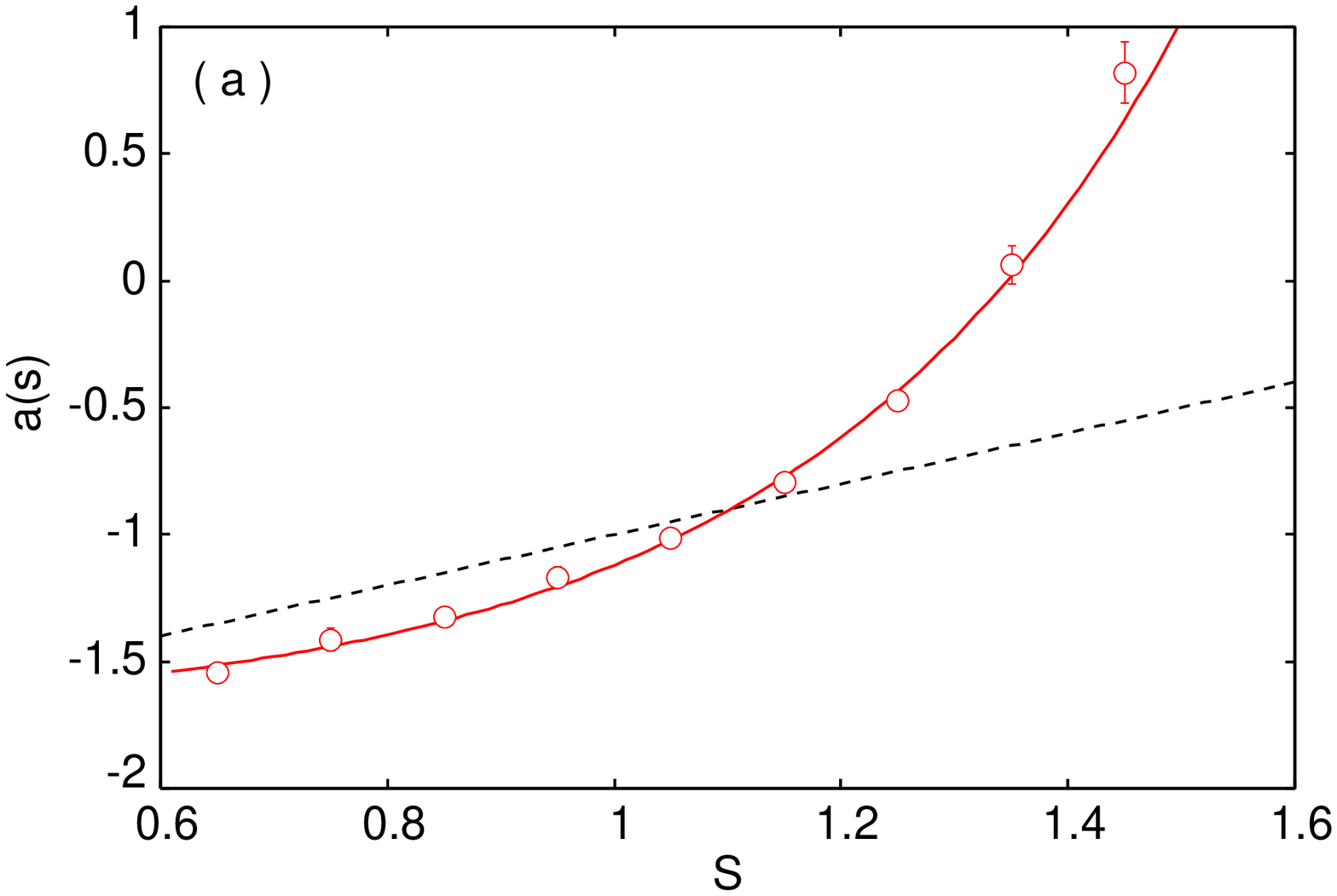}
\end{minipage}
\begin{minipage}[b]{8.0cm}
\includegraphics*[width=7.5cm]{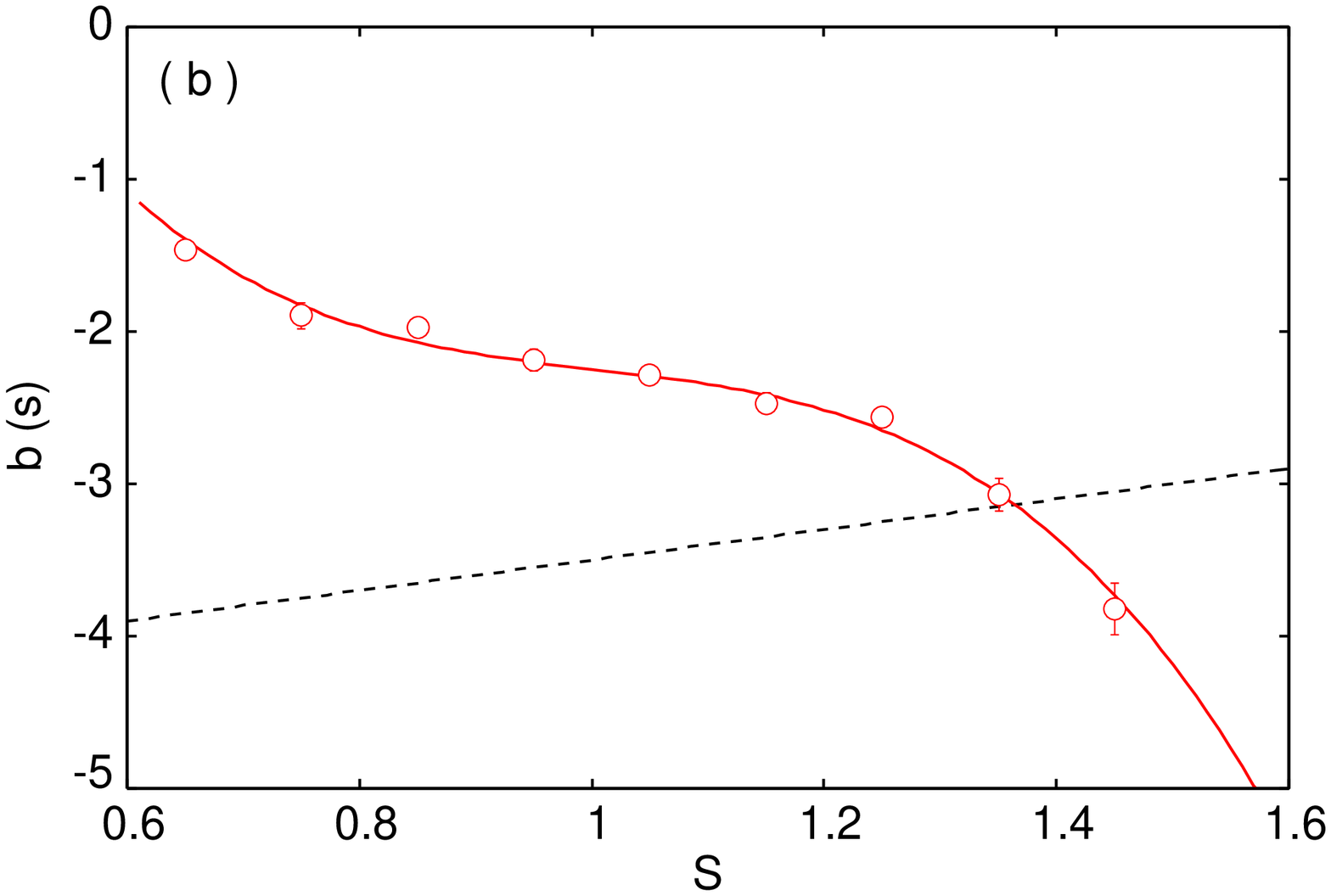}
\end{minipage}
\end{center}
\caption{The numerical values of  $a$ and $b$ are plotted
as a function of $s$, where original definitions
of $a(s)$ and $b(s)$ in NKG function are shown by the dotted lines,
and the open circles denote the averaged MC data which are
fitted by empirical formulae shown by red lines.  See the text.
}
\label{fig:2}
\end{figure}

The determination of the lateral distribution function of an air shower
is substantially important in this experiment,
since the total number of shower particles is obtained by 
fitting the structure function to the experimental data.
Since our air-shower array is also used to study the TeV gamma-ray point
sources, lead plate of 5 mm thick is placed
above each scintillation counter to increase the sensitivity,
the detector response should be carefully examined by MC.
Using the Monte Carlo data obtained under the same conditions as the experiment,
we find that the following modified NKG function can fit
well to the lateral distribution of shower particles under the lead
plate:

\begin{equation}
\label{NKG_func}
f(r,s) = \frac{N_e}{C(s)} ({\frac{r}{{r_{m}}^{'}}})^{a(s)}{(1+{\frac{r}{{r_m}^{'}}})}^{b(s)} /{r_{m}}^{'2}
\end{equation}

\begin{equation}
C(s)=2\pi B(a(s) + 2,-b(s)-a(s)-2)
\end{equation}

where ${r_{m}}^{'}$ = 30 m, 
and the variable $s$ corresponds to the age parameter, $N_e$ the total number
of shower particles and $B$ denotes the beta function. 
The functions $a(s)$ and $b(s)$ are determined as follows.
In CORSIKA simulation, the shower age parameter $s$ is calculated at
observation level by fitting to a function for the one dimensional 
shower development. It may be possible to assume that air showers with the same shower age s
are in the almost same stage of air shower development in the atmosphere, i.e. 
they show the almost same lateral distribution for shower particles irrespective of 
their primary energies. The lateral distribution of the particle density
obtained by the simulation with carpet array configuration 
is normalized by the total number of particles which is derived from
the total energy deposit in infinitely wide scintillator.
These events are then classified according to the stage of
air shower development using the age parameter
and they are averaged over the classified events. The fitting of the
equation (\ref{NKG_func}) to the averaged MC data is made to
obtain the numerical values $a$ and $b$. Thus, we can obtain the behavior of
$a$ and $b$ as a function of $s$ as shown in Fig.~\ref{fig:2} 
where original definitions of $a(s)$ and $b(s)$ in NKG function 
are shown by the dotted lines. Although our result shows different dependences
of $a$ and $b$ on $s$, it is confirmed that the lateral distribution of
the shower particles is better reproduced by our formula.
This expression is valid in the range of s = 0.6 $\sim$ 1.6, 
$\sec\theta < $ 1.1 and r = 5 $\sim$ 3000 m. 

%
% Fig.2(a)
%
%
%
% Fig.2(b)
%
%

\begin{figure}[htbp]
\begin{center}
\begin{minipage}[b]{8.0cm}
\includegraphics*[width=7.5cm]{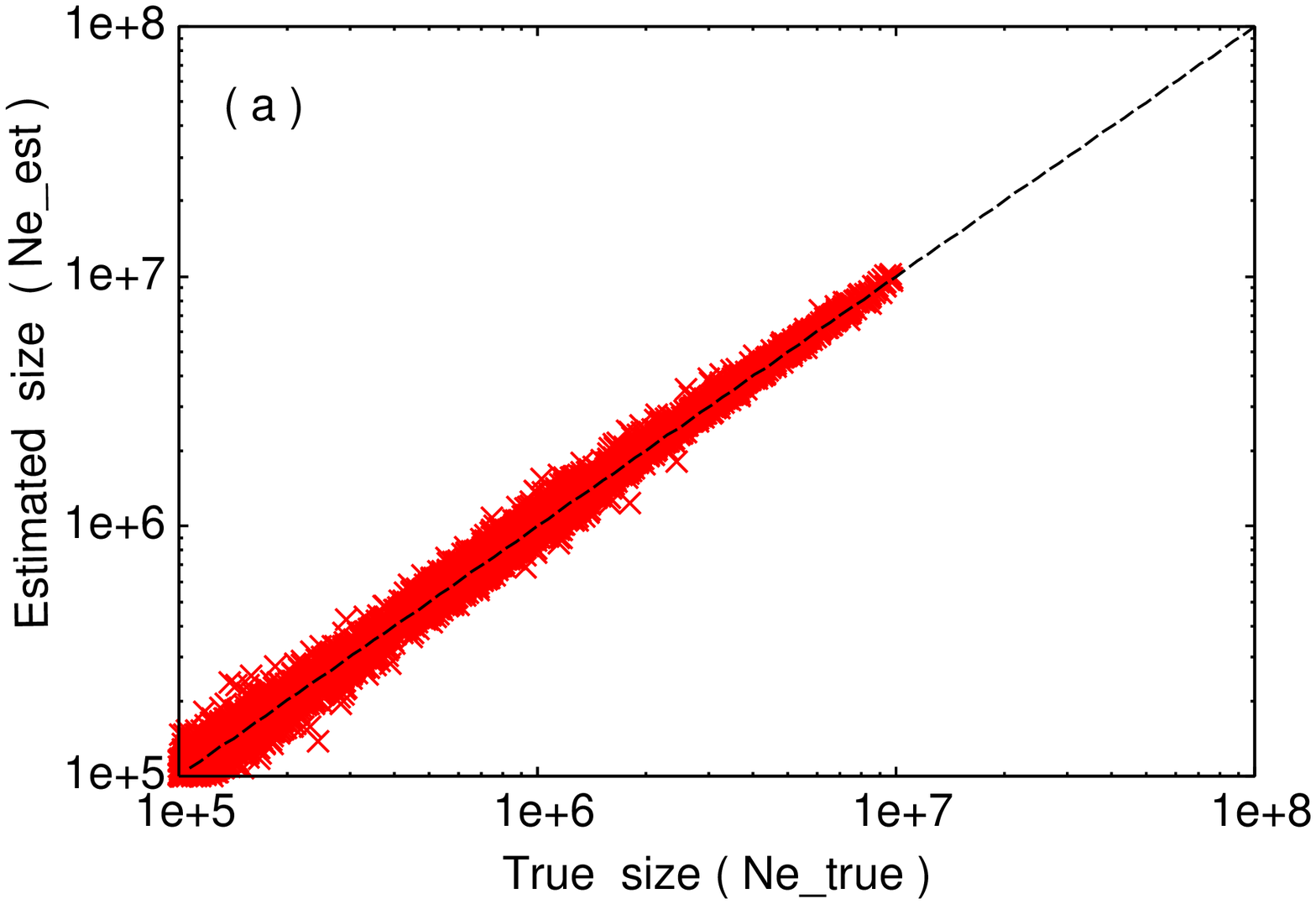}
\end{minipage}
\begin{minipage}[b]{8.0cm}
\includegraphics*[width=7.5cm]{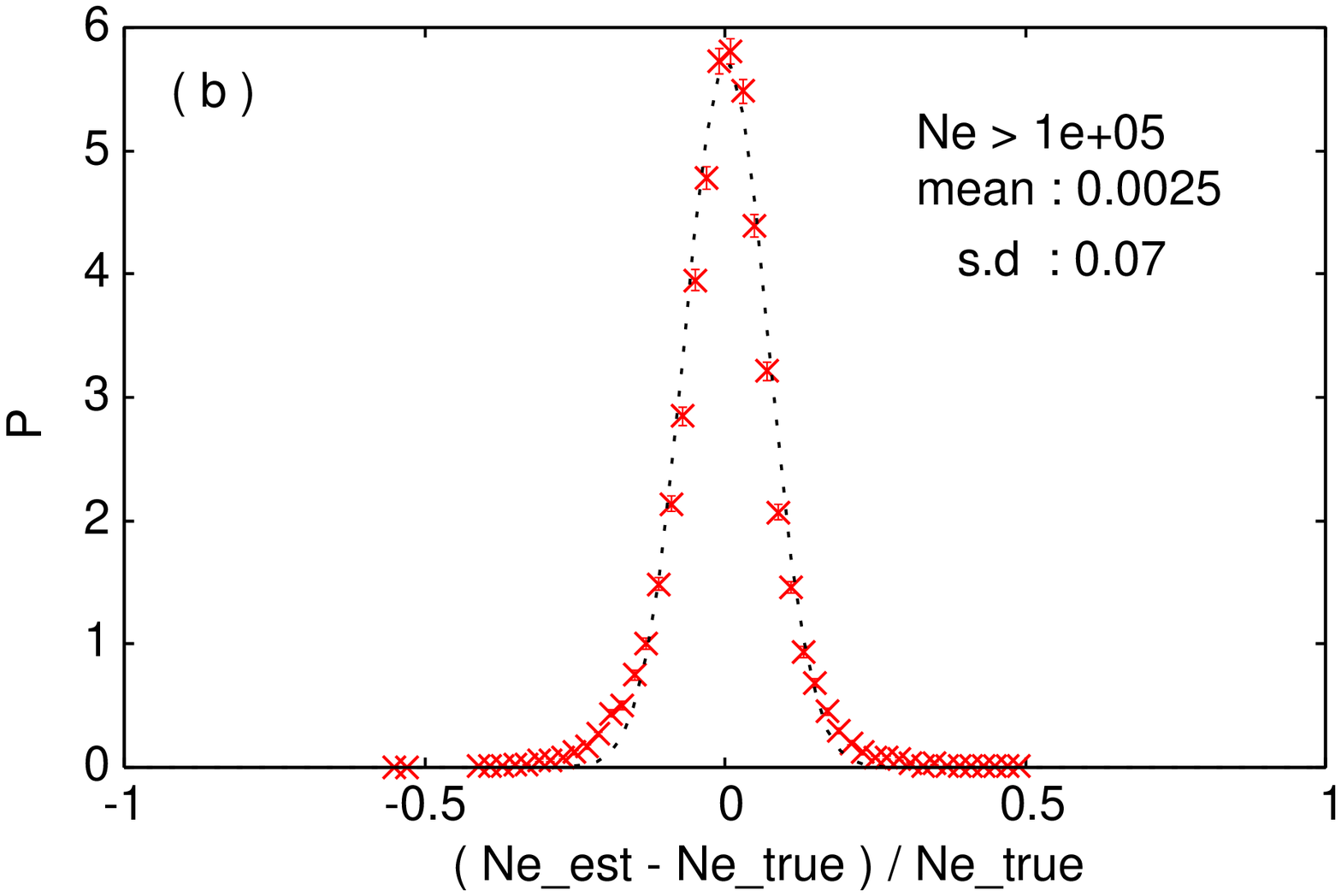}
\end{minipage}
\end{center}
\caption{(a) The correlation between the true shower size ($Ne_{true}$) and the 
estimated shower size ($Ne_{est}$). (b) The shower size resolution is 
estimated to be 7\% above  $Ne >$ $10^{5}$ with $\sec\theta$ $\leq$ 1.1. 
}
\label{fig:3}
\end{figure}

Based on the Monte Carlo simulation, the correlation between the true shower 
size and the estimated shower size is demonstrated in Fig.~\ref{fig:3}(a) and 
Fig.~\ref{fig:3}(b). As seen in these two figures, a good correlation between
the true shower size and
the estimated shower size is obtained and the shower size resolution is 
estimated to be 7\% above $N_e$ $\geq$ $10^{5}$ with $\sec\theta$ $\leq$ 1.1. 
 
%
% Fig.3(a)
%
%
%
% Fig.3(b)
%
%

\subsection{Relation between the shower size and the primary energy}
%
% Fig.4
%
%

\begin{figure}[t]
\begin{center}
\includegraphics*[width=7.5cm]{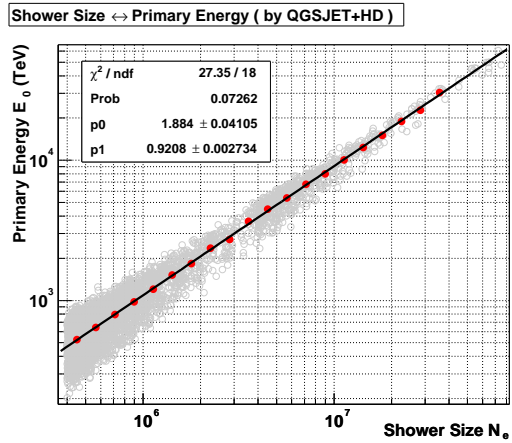}
\end{center}
\caption{Correlations of the air-shower size $N_e$ and the primary energy $E_{0}$ 
at the Tibet observation level for the QGSJET+HD model with $\sec\theta$ $\leq$ 1.1.
The solid circles denote the average values, and the solid lines is a fit by equation ( 3 )}
\label{fig:4}
\end{figure}

In Fig.~\ref{fig:4}, we show the correlation between the shower size $N_e$ 
under the lead plate at the Tibet observation level 
and the primary energy $E_0$ 
for the QGSJET+HD model. The conversion function from the shower size $N_e$ 
to the primary energy $E_0$ can be expressed by the following equation 
for $\sec\theta$ $\leq$ 1.1,  

\begin{equation}
E_0 = a \times {(\frac{N_e}{1000})}^{b} \mbox{\hspace{2cm}  [TeV]},
\end{equation}

where a=1.88, b=0.92.
The energy resolution is estimated by our simulation as 36\% and 17\% at energies 
around 200 TeV and 2000 TeV, respectively.
It is seen that the energy estimation of the primary particle from the air shower size is
well made with a good accuracy in the wide energy range including the knee region. 
We further examine the dependence of the result on the interaction models, primary models,
and  other possible systematic biases in the very near future.

%%%%%%%%%%%%%%%%%%%%%%%%%%%%%%%%%%%%%%%%%%%%%%%%%%%%%%%%%%%%%%%%%%%%%%%
\section{Results and Discussions}

\begin{figure}[t]
\begin{center}
\includegraphics*[width=7.5cm]{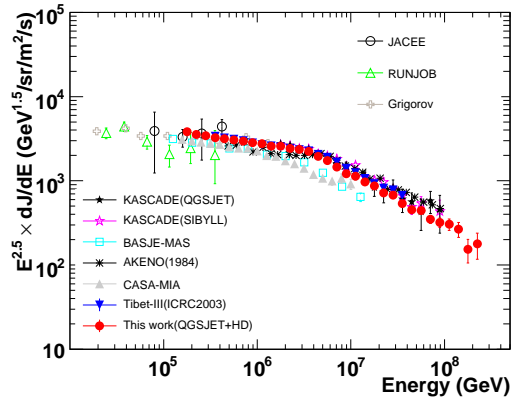}
\end{center}
\caption{Differential all-particle cosmic-ray flux in a wide range over 3 decades between $10^{14}$ eV  
and $10^{17}$ eV measured by the Tibet-III air-shower array using the QGSJET+HD model. This work 
is compared with other experiments :
(JACEE \cite{Asakimori1}, RUNJOB \cite{Apanasenko1}, PROTON satellite \cite{Grigorov}, 
KASCADE \cite{Antoni}, BASJE-MAS \cite{Ogio}, AKENO \cite{Nagano1}, CASA-MIA \cite{Glasmacher}, 
Tibet-III \cite{Amenomori4}).
}
\label{fig:5}
\end{figure}

 We obtained the all-particle 
energy  spectrum between 1 $\times$ $10^{14}$ eV and 1 $\times$ $10^{17}$ eV 
based on the QGSJET+HD model as shown in Fig.~\ref{fig:5}. The red circle 
represents this work, and the blue inverse-triangle is our 
old result \cite{Amenomori4}, and they are compared with other experiments. 
Our new result shows a good agreement  with the previous one in the energy range less than 4
PeV, while suggesting a slightly steeper power index  above 4 PeV and 
about 10\% lower intensity around $10^{16}$ eV. This difference may be due to 
the upgrade of  MC calculations and the increase of the observed data. 
The spectral index of this work is -2.68 $\pm$ 0.02(stat.) below 1 PeV and  -3.12 $\pm$ 0.01(stat.) 
above 4 PeV in the case of QGSJET+HD model. The evaluation of various systematic errors 
is  under examination at present.
 
%
% Fig.5
%
%
The all-particle spectrum obtained in a wide range over 3 decades between $10^{14}$ eV  
and $10^{17}$ eV clearly shows the position of the knee being at energy around 4 PeV. 
In this work, we obtained the result based on the assumption that the knee energy region
is dominated by nuclei heavier than helium \cite{Amenomori5}. It is noted that the
all-particle spectrum depends slightly on the primary composition in the energy
region below about $5 \times 10^{14}$ eV. A further study of the dependence of the spectrum  on  
 the simulation codes, interaction models, primary composition models, and other possible 
systematic biases should  be carefully done in the very near future.

\section*{Acknowledgments}
The collaborative experiment of the Tibet Air Shower Arrays has been
performed under the auspices of the Ministry of Science and Technology
of China and the Ministry of Foreign Affairs of Japan. This work
was supported in part by Grants-in-Aid for Scientific Research
on Priority Areas (712) (MEXT), by Scientific Research (JSPS) in
Japan, by the National Natural Science Foundation of China, and by
the Chinese Academy of Sciences.

%%%%%%%%%%%%%%%%%%%%%%%%%%%%%%%%%%%%%%%%%%%%%%%%%%%%%%%%%%%%%%%%%%%%%%%

\clearpage

%
% Fig.1
%
%

%
% Fig.2(a) and 2(b)
%

%
% Fig.3(a) and 3(b)
%

%
% Fig.4
%
%

%
% Fig.5
%

\end{document}